\documentclass[twocolumn,superscriptaddress,prl]{revtex4-1}
\usepackage{mathrsfs,braket}
\usepackage{amssymb, amsbsy, amsmath, latexsym, dsfont, array, layout,
graphicx,mathrsfs,color,ulem,bm}
\DeclareMathOperator{\sign}{sign}
\usepackage{changes}

\begin{document}

\title{Molecular chaos in dense active systems}

\author{Lu Chen}
\thanks{These authors contributed equally to this work}
\affiliation{Department of Physics, Beijing Normal University, Beijing 100875, China}
\affiliation{Complex Systems Division, Beijing Computational Science Research Center, Beijing 100193, China}
\author{Kyle J. Welch}
\thanks{These authors contributed equally to this work}
\affiliation{Complex Systems Division, Beijing Computational Science Research Center, Beijing 100193, China}
\affiliation{Department of Chemical Engineering and Materials Science, University of Minnesota, Minneapolis, MN 55455, USA}
\author{Premkumar Leishangthem}
\thanks{These authors contributed equally to this work}
\affiliation{Complex Systems Division, Beijing Computational Science Research Center, Beijing 100193, China}
\author{Dipanjan Ghosh}
\affiliation{Department of Chemical Engineering and Materials Science, University of Minnesota, Minneapolis, MN 55455, USA}
\author{Bokai Zhang}
\affiliation{Complex Systems Division, Beijing Computational Science Research Center, Beijing 100193, China}
\author{Ting-Pi Sun}
\affiliation{Department of Chemical Engineering and Materials Science, University of Minnesota, Minneapolis, MN 55455, USA}
\author{Josh Klukas}
\affiliation{Department of Chemical Engineering and Materials Science, University of Minnesota, Minneapolis, MN 55455, USA}
\author{Zhanchun Tu}
\affiliation{Department of Physics, Beijing Normal University, Beijing 100875, China}
\author{Xiang Cheng}\email{xcheng@umn.edu}
\affiliation{Department of Chemical Engineering and Materials Science, University of Minnesota, Minneapolis, MN 55455, USA}
\author{Xinliang Xu}\email{xinliang@csrc.ac.cn}
\affiliation{Complex Systems Division, Beijing Computational Science Research Center, Beijing 100193, China}

\begin{abstract}
{\bf
The hypothesis of molecular chaos plays the central role in kinetic theory, which provides a closure leading to the Boltzmann equation for quantitative description of classic fluids. Yet how to properly extend it to active systems is still an open question in nonequilibrium physics. Combining experiment, simulation, and theory, we investigate the emergent collective behaviors of self-propelled particles that exhibit collision avoidance, a moving strategy commonly adopted in natural and engineering active systems. This dense active system shows unusual phase dynamics strongly regulated by many-body interactions, which cannot be explained by theories assuming molecular chaos. To rationalize the interplay between different emergent phases, a simple kinetic model is proposed with a revised molecular chaos hypothesis, which treats the many-body effect implicitly via categorizing different types of particle pair collisions. Our model predicts an optimal growth rate of flocking and illustrates a generic approach for understanding dense active systems.
}
\end{abstract}

\maketitle

From swarming bacteria \cite{Dunkel_2013,Guo_2018,Liu_2021}, to flocking birds \cite{Nagy_2010,Bialek_2012}, and even human crowds \cite{Karamouzas_2014,Murakami_2021}, active systems of self-propelled individuals exhibit emergent behaviors that are both fascinating and perplexing \cite{Marchetti_2013,Chate_2020,Bar_2020}. For systems of passive particles, the Boltzmann equation obtained by assuming molecular chaos provides a simple yet powerful tool elucidating the emergence of macroscopic properties from microscopic particle interactions. For active systems, the molecular chaos hypothesis is widely used in kinetic approaches such as the Smoluchowski \cite{Baskaran_2010} or the Boltzmann approaches \cite{Bertin_2009, Peshkov_2014, Patelli_boltzmann_2019}, and has been proven successful in explaining observations for dilute systems \cite{Bertin_2006, Lam_polar_2015, Lam_selfpropelled_2015, Denk_2020, Ihle_2011}. However, recent experiments \cite{Suzuki_2015,Ghosh_2022} and simulations \cite{Thuroff_2013} have shown that such a simple hypothesis is inadequate for even moderately dense active systems. While dense active systems can be qualitatively described by coarse-grained theories with \textit{ad hoc} transport coefficients \cite{Geyer_2019, Grobmann_2020}, how to derive these coefficients from microscopic interactions remains elusive. To address this challenge, various approaches have been developed, all of which involve a complicated hierarchy of equations solved by truncation \cite{Chou_2015, Patelli_landau_2021}. Contrary to these previous approaches, for dense active systems with clusters we identify the crucial many-body effect that leads to dynamics qualitatively different from the molecular chaos prediction. A simple kinetic model is proposed, which avoids complex hierarchy schemes by capturing the many-body effect implicitly so that molecular chaos resumes applicable at high densities in a modified form.

Our active system is composed of self-propelled particles that avoid collision via active reorientation. Such a collision avoidance behavior is commonly adopted by individuals in cell colonies \cite{Smeets_2016}, animal herds \cite{Chan_2013}, swarming robots \cite{Rubenstein_2014,Aguilar_2018}, and traffic flows \cite{Helbing_2001,Ou_2018}. Unlike the classic Vicsek model \cite{Vicsek_1995} that imposes an explicit local alignment rule, collision avoidance dictates particle interaction in our system, where the local alignment rule manifests itself as a coarse-grained average of particle dynamics. The coupling between self-propulsion and active reorientation gives rise to rich phase dynamics, which exhibit a complex interplay between motility-induced phase separation (MIPS) \cite{Palacci_2013,Redner_2013,Cates_2015,Turci_2021} and the formation of polarly-aligned flocks \cite{Toner_1995,Bricard_2013,Solon_2015}. While dynamic interplay between different emergent phases has been frequently observed as a generic feature of active systems with orientation-dictated interactions \cite{Zhang_2021,Linden_2019,Grobmann_2020,Sansa_2021}, how such interplay arises from microscopic interactions is still far from understood. Thanks to the simplicity of our system and the new theoretical approach adopted, we quantitatively analyze MIPS and flocking, and conclude their interplay is well described by an analytic relation between the transport coefficient of flocking and the fraction of particles in clusters. As such, our results provide not only a detailed understanding of the dynamics of dense active systems with a ubiquitously-observed particle interaction rule but also a design principle for active robotic systems optimizing the rise of collective dynamics \cite{Douglas_2012,Rubenstein_2014,Aguilar_2018}.

\parskip = 5pt plus 1pt

\begin{figure}[t]
   \includegraphics[width=0.46\textwidth]{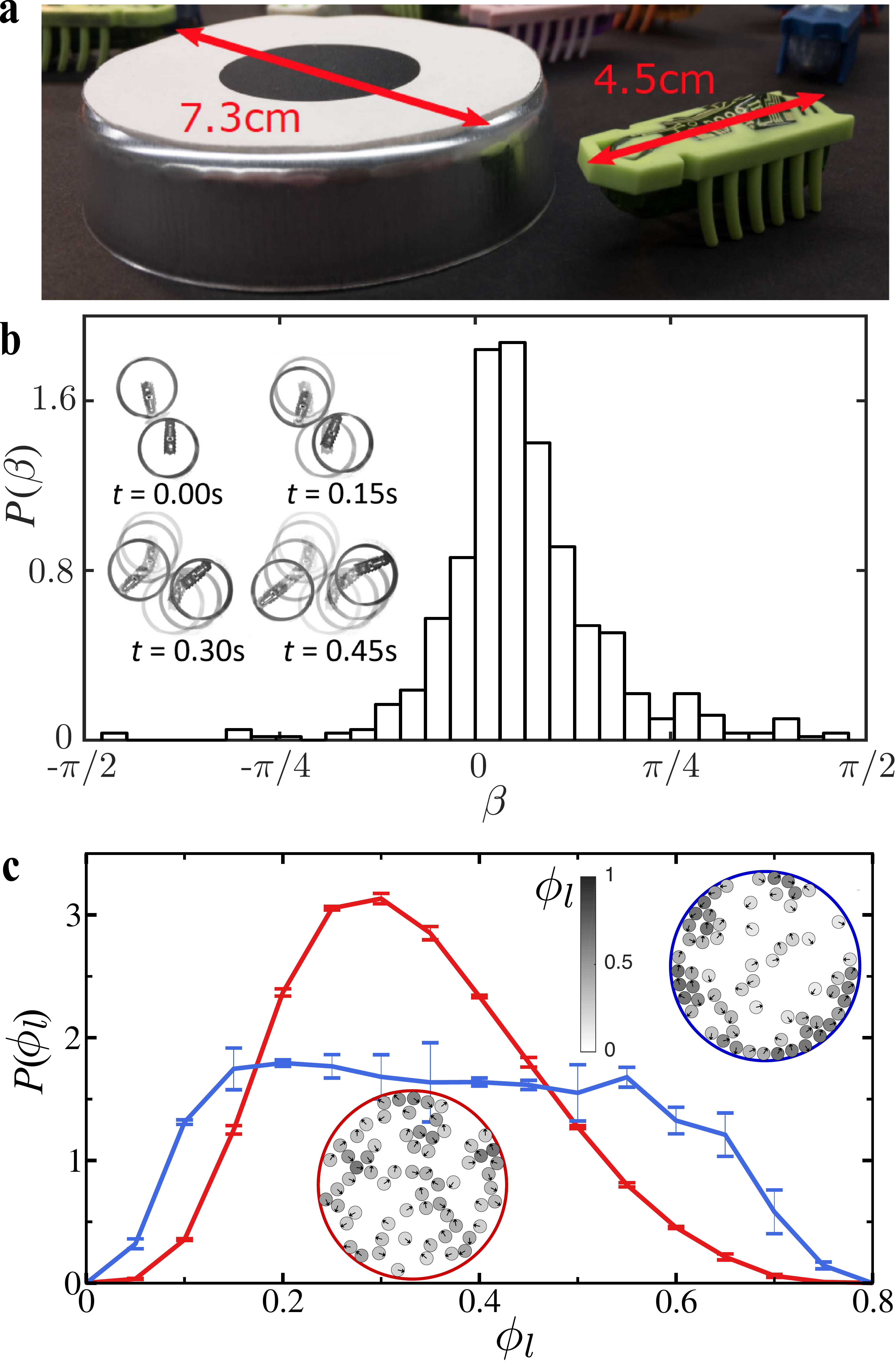}
	\caption{{\bf Experiments.} {\bf a}, An active particle is composed of aluminum dish (left) over a hexbug (right). {\bf b}, Histogram of $\beta$, the turning angle per frame for particles during collision. Inset: a typical pair collision. {\bf c}, Histogram of the local density $\phi_l$ for systems of 65 particles with (red line) and without (blue line) active reorientation. Inset: images of the system with (left) and without (right) active reorientation. The gray-scale of each particle represents the local density of the particle $\phi_l$.}\label{fig:Fig1}
\end{figure}

\noindent{\bf Experiment}
\parskip = 0pt plus 1pt

We construct a table-top active granular experiment utilizing toy vibrobots (Hexbug Nano) \cite{Li_2013,Deblais_2018}. The hexbugs are placed under overturned circular weighing boats of diameter $d=7.3$ cm (Fig.\,\ref{fig:Fig1}a). An active particle thus consists of a weighing boat powered by an internal vibrobot, which exhibits self-propulsion as well as random rotation. The design gives rise to unique particle dynamics during collision, different from previous experiments using vibrated disks with build-in polarity \cite{Deseigne_2010}. When two active particles collide in our system, the boats rotate around the point of contact due to the torque exerted by the hexbugs, which lead to the reorientation of the hexbugs within the boats turning away from the contact point (Fig.\,\ref{fig:Fig1}b inset). As a result, the two particles tend to avoid further collisions. The histogram of the turning angle for each particle during collision shows a positive mean of $0.2$, indicating a clear trend of collision avoidance (Fig.\,\ref{fig:Fig1}b).

The experiment is conducted with $N$ particles confined in a circular arena of a diameter of 1 m, where $N$ ranges from 5 (area fraction $\phi =0.03$) to 140 ($\phi =0.75$). For each $\phi$, two independent runs, each 3000 frames, are recorded at 30 Hz. The root-mean-square velocity of particles at short times decreases linearly with increasing $\phi$ (Supplementary Information (SI) Sec. A), a feature normally leading to MIPS for active Brownian particles \cite{Redner_2013}. However, phase separation is suppressed by active reorientation in our system at even high densities: for a system of 65 particles ($\phi = 0.3$), the distribution of local density of each particle $\phi_l$, i.e., the ratio of particle size to the size of its Voronoi cell, shows one well-defined peak near the system average $\phi$ (Fig.\,\ref{fig:Fig1}c). In comparison, for 65 particles with active reorientation turned off by refraining hexbugs from rotation inside boats, the local density distribution broadens significantly with a notable portion of particles in dense clusters with $\phi_l>0.6$ (Fig.\,\ref{fig:Fig1}c). Further analysis of particle dynamics shall be presented through a quantitative comparison with theoretical and numerical findings in due course. 

\parskip = 5pt plus 1pt

\begin{figure}
	\includegraphics[width=0.46\textwidth]{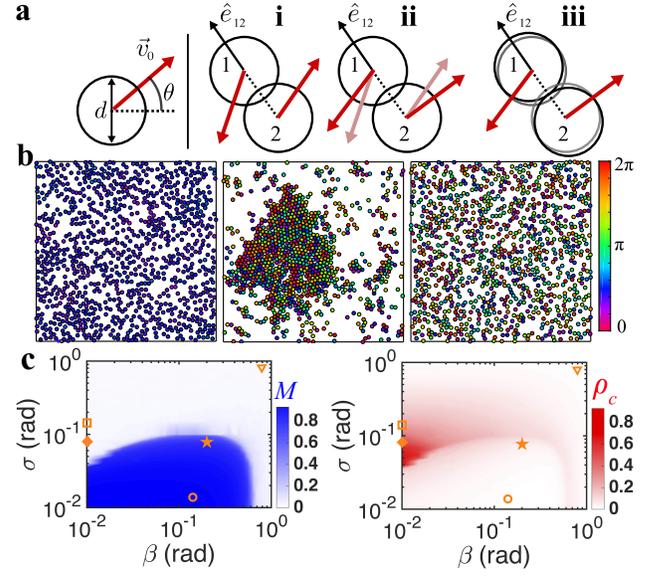}
    \caption{{\bf Simulations.} {\bf a}, Active reorientation during a collision. {\bf b}, Flocking (left), clustering (center) and disordered (right) phases, where the color of each particle denotes its orientation $\theta$. Phase diagram of the steady-state flocking order parameter $M$ ({\bf c}) and clustering parameter $\rho_{c}$ ({\bf d}) in a phase space defined by active reorientation $\beta$ and rotational noise $\sigma$. The examples in {\bf b} and our experiments are marked in the phase space: flocking (circle), clustering (square), disorder (triangle) from simulations, and experiments with (star) and without (diamond) active reorientation.}\label{fig:Fig2}
\end{figure}

\noindent{\bf Simulation}
\parskip = 0pt plus 1pt

We conduct a numerical study of self-propelled Brownian particles performing active reorientation (Fig.\,\ref{fig:Fig2}a). The equation of motion for particle $i$ at location $\vec{r}_i$ with orientation $\vec{n}_i\equiv(\cos \theta_i,\sin \theta_i)$ is given as: 
\begin{equation}
\frac{d\vec{r}_i}{dt}=v_0 \vec{n}_i \prod_{j\neq i}^{N} (1-\epsilon_{ij}) + \frac{1}{\xi_t}\sum_{j\neq i}^{N} \vec{F}_{ji},
\label{translational-motion}
\end{equation}
\begin{equation}
\frac{d\theta_i}{dt}=\sum_{j\neq i}^{N} \beta \epsilon_{ij} \sign (\theta_{ji}-\theta_{i}) + \sigma \eta_{i}(t).
\label{rotational-motion}
\end{equation}
Here, $v_0$ is the self-propulsion speed, $\xi_t$ is the translational friction coefficient, $\sigma = \sqrt{2D_r}$ with $D_r$ denoting the rotational diffusion coefficient and $\eta_{i}$ is a Gaussian white noise with zero mean and unit variance. We define collision as a state of overlap between two particles and set $\epsilon_{ij}=1$ (or $\epsilon_{ij}=0$) if particle $i$ overlaps with $j$ (or not). The inter-particle repulsion $\vec{F}_{ji}$ applies only during collision as $\vec{F}_{ji}=\epsilon_{ij} k(d-|\vec{r}_{ij}|)\hat{e}_{ij}$, where $d$ is particle diameter, $\vec{r}_{ij} \equiv \vec{r}_i - \vec{r}_j$, $\hat{e}_{ij}$ is the unit vector along $\vec{r}_{ij}$ with $\hat{e}_{ij}\equiv\vec{r}_{ij}/|r_{ij}| = (\cos \theta_{ij},\sin \theta_{ij})$, and the elastic constant $k$ is chosen so that colliding particles are pushed apart and the overlap is removed in one time step. Soft repulsive potentials $\sim r_{ij}^{-n}$ with $n=4$ or $6$ have also been tested, which yield quantitatively similar results. Note that the self-propulsion speed of a colliding particle is zero as the hexbug rotates inside the boat without pushing forward. Lastly, collision-avoiding active reorientation is implemented by the rotation of $\vec{n}_{i}$ towards $\hat{e}_{ij}$ during collision with an angular speed $\beta$ (Fig.\,\ref{fig:Fig2}a). Such an active reorientation rule has also been used to simulate the contact inhibition of cell locomotion \cite{Smeets_2016}. 

\begin{figure*}[t]
	\includegraphics[width=0.8\textwidth]{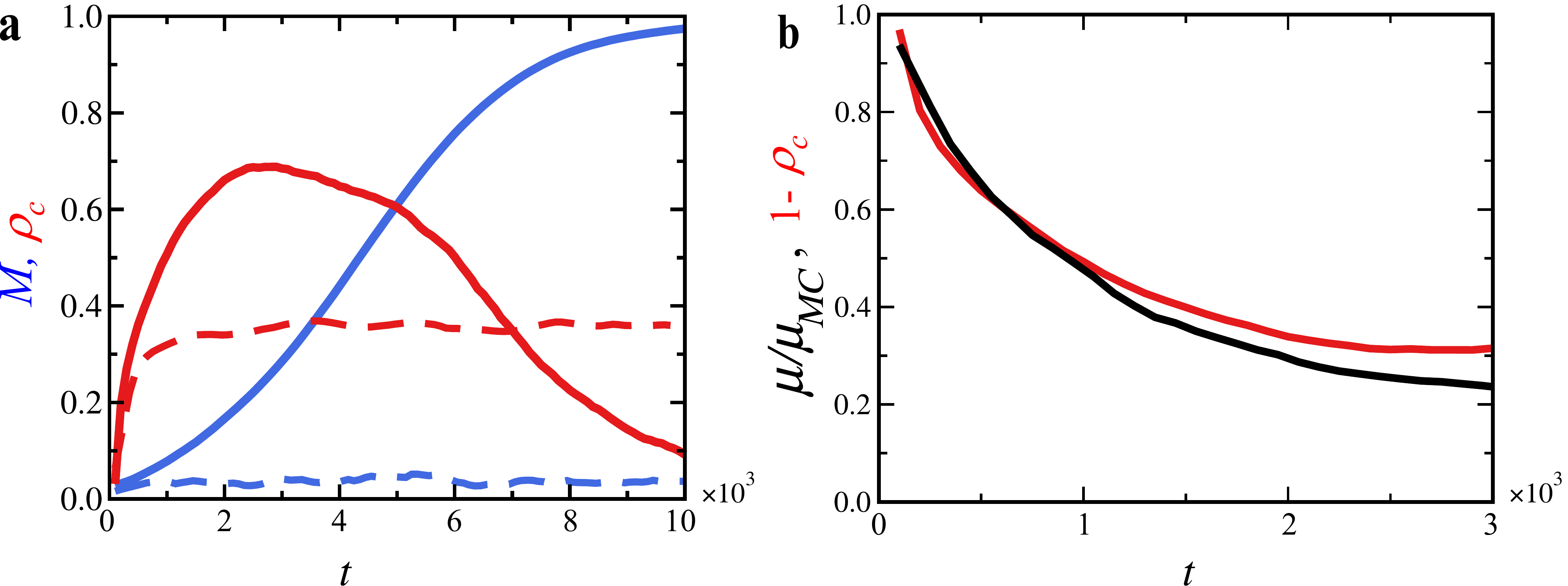}
    \caption{{\bf System dynamics from simulations.} {\bf a}, The temporal evolution of flocking $M$ (blue line) and clustering $\rho_{c}$ (red line) for systems with $\sigma=\beta=0.01$ (solid line), and $\sigma=0.1$ and $\beta=0.01$ (dashed line). {\bf b}, The temporal evolution of the transport coefficient $\mu$ (black line) and $1-\rho_c$ (red line) with $\sigma=\beta=0.01$, where $\mu$ is normalized by the molecular chaos prediction $\mu_{MC}$.}\label{fig:Fig3}
\end{figure*}

We use particle diameter and simulation time step as the unit of length and time. In each simulation $N$ particles are placed in a square box of side length $L$ with random locations and orientations. A periodic boundary condition is adopted for both dimensions. We fix $N=1000$, where the finite-size effect becomes negligible (SI Sec.\,B).

The dynamics of our system are controlled by four parameters: i.e., $v_0$, $\beta$, $\sigma$ and $\phi$. When using the experimentally extracted values $v_0=0.1$, $\beta=0.2$ and $\sigma=0.08$, our simulation quantitatively reproduces the mean squared displacements of particles at different $\phi$ (Fig.\,S1c). While our experiments only allow two choices of $\beta$, being 0.2 (or 0) with active reorientation on (or off), in simulation we fix $v_0=0.1$ and vary $\sigma$ and $\beta$ systematically to explore system dynamics over a larger parameter space. For each pair of $\{\sigma,\beta\}$, we perform 100 independent  runs to obtain the average. Finally, unless stated otherwise, we fix $\phi=0.3$ for a quantitative comparison between theory and experiment, where the system exhibits rich phase dynamics with distinct disordered, clustering and flocking phases (Fig.\,\ref{fig:Fig2}b).

The average orientation of particles is quantified by $\vec{M}=\langle\vec{n}\rangle_{N}$, where the average $\langle \cdot \rangle_N$ is taken over all $N$ particles. $M=|\vec{M}|$ ranges between zero for a disordered phase and one for a perfectly aligned flocking phase. As illustrated by the steady state phase diagram of $M$ (Fig.\,\ref{fig:Fig2}c), our experiments with $\sigma=0.08$ and $\beta=0.2$ fall within a flocking phase arise from active reorientation. Without active reorientation ($\beta = 0$), no flocking phase is observed (SI Sec.\,G).

MIPS is characterized by the distribution of local density $\phi_l$ \cite{Redner_2013}: $P(\phi_l)$ has a single peak around the system average $\phi$ when the system is in a single phase, whereas $P(\phi_l)$ broadens and splits into two separate peaks as the system phase separates (Fig.\,S3). We quantify the degree of MIPS by the fraction of particles in dense clusters, $\rho_{c} \in [0, 1]$. Here, $\rho_c$ is taken as the fraction of particles with $\phi_l>0.6$, while more sophisticated methods in estimating $\rho_c$ yield quantitatively similar results (SI Sec.\,C). Figure\,\ref{fig:Fig2}d shows the steady state phase diagram of $\rho_c$. Without active reorientation, our results are quantitatively similar to those from previous studies of MIPS (Fig.\,S4) \cite{Redner_2013}. As expected, our experiments without active reorientation fall in the MIPS regime (Fig.\,\ref{fig:Fig2}d).

\parskip = 5pt plus 1pt

\noindent{\bf Interplay between flocking and MIPS}
\parskip = 0pt plus 1pt

The complementary shapes of the steady state phase diagram of $M$ and $\rho_{c}$ indicate the mutually exclusive nature of flocking and MIPS (Figs.\,\ref{fig:Fig2}c and 2d), a trend that becomes more evident in the transient kinetics. For systems with steady state flocking (e.g., $\sigma=\beta=0.01$), $\rho_{c}$ increases substantially with a clear MIPS at short times when $M$ is small. However, this transient presence of MIPS collapses at long times when flocking is fully developed (Fig.\,\ref{fig:Fig3}a). In contrast, for systems with steady state MIPS (e.g., $\sigma=0.1$ and $\beta=0.01$), $\rho_{c}$ reaches a high plateau and remains stable (Fig.\,\ref{fig:Fig3}a), whereas flocking never develops. The suppression of MIPS by flocking is in agreement with previous studies on MIPS that arises from collision-induced slowdown at high densities \cite{Peruani_2011, Shi_2018}: orientation alignment in the flocking state drastically reduces the number of collisions and therefore eliminates the driven factor underlying MIPS.

Conversely, clustering shows a more subtle influence on the growth of flocking, which is qualitatively different from the molecular chaos prediction. Following the generalized Boltzmann approach \cite{Bertin_2009, Peshkov_2014}, the kinetic equation for $M$ has been obtained in the linear order of $M$ \cite{Lam_polar_2015,Lam_selfpropelled_2015} (SI Sec.\,D)
\begin{equation}
\frac{1}{M}\frac{dM}{dt}=\lambda\mu-\frac{\sigma^2}{2},
\label{kinetic-theory}
\end{equation} 
where $\lambda$ is the collision rate and $\sigma^2/2 = D_r$ characterizes the effect of self-diffusion. The transport coefficient, $\mu \equiv \langle\vec{p}\cdot\delta\vec{p}\rangle$, measures the momentum change projected onto the alignment direction in each collision, where $\vec{p}=\vec{n}_i+\vec{n}_j$ is the total momentum of the two colliding particles before the collision and $\delta\vec{p}$ is the change of $\vec{p}$ due to the collision. The average $\langle \cdot \rangle$ is taken over all collisions occurred during the $dt$ under consideration.

In Eq.\,\ref{kinetic-theory}, $\lambda$ and $M$ are static quantities as they can be obtained from a static configuration, while $\mu$ is a dynamic quantity that requires knowledge about microscopic dynamics during collision. To bridge the timescale of microscopic dynamics and that of macroscopic properties including the global polar alignment $M$, the essence of a kinetic model is to predict the transport coefficient from microscopic interactions. Therefore, we focus here on the influence of clustering on $\mu$. Under the assumption of molecular chaos, binary collisions are statistically independent and all the collision configurations are equally probable. The assumption leads to a constant $\mu \equiv \mu_{MC} = 2\beta/\pi$, independent of the degree of clustering $\rho_c(t)$ (SI Sec.\,E). To verify the prediction, we measure $\mu(t)$ in simulations using Eq.\,\ref{kinetic-theory}, which by setting $dt = 1$ manifests in the form of $\mu=\lambda^{-1}(dM/M+\sigma^2/2)$, where $\lambda$, $M$ and $dM$ are obtained from static configurations. Figure~\ref{fig:Fig3}b shows $\mu(t)$ for $\beta=\sigma=0.01$, which is a decreasing function qualitatively different from the constant predicted by molecular chaos. Note that we only analyze data for $t<3000$ where $M<0.3$, so that the use of the linear order equation (Eq.\,\ref{kinetic-theory}) is justified.

\parskip = 5pt plus 1pt

\begin{figure}[t]
	\includegraphics[width=0.46\textwidth]{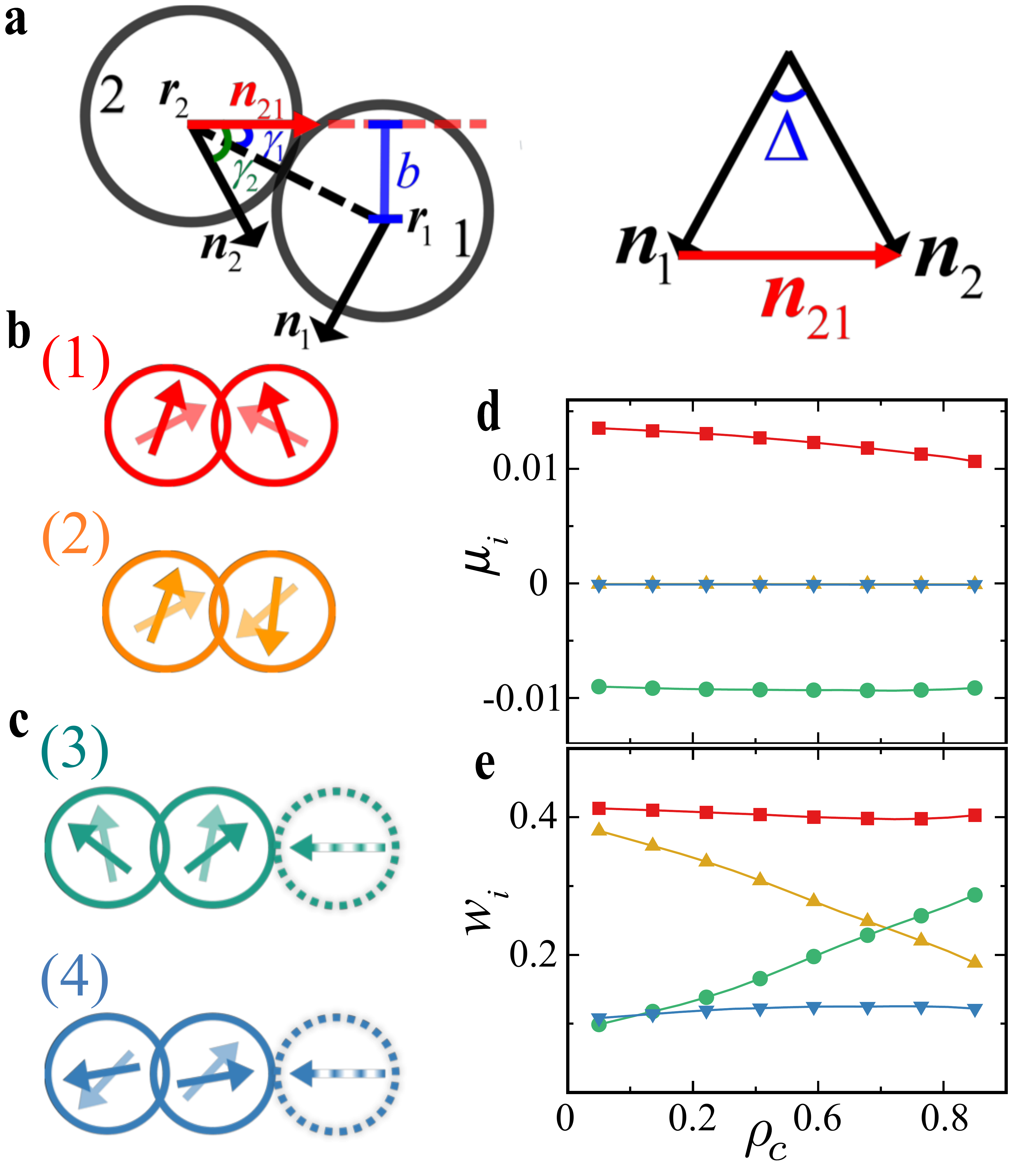}
    \caption{{\bf Categorization of binary collisions.} {\bf a}, Parameters for a generic binary collision event. Examples of normal collisions ({\bf b}) and abnormal collisions ({\bf c}): (1) Normal and aligned, (2) Normal and anti-aligned, (3) Abnormal and aligned, and (4) Abnormal and anti-aligned. The light (dark) colored arrows indicate the incoming (outgoing) orientations. Neighboring particles that affect the binary collisions are shown as dashed circles. The average momentum change $\mu_i$ ({\bf d}) and the relative frequency $w_i$ ({\bf e}) for each category are obtained from simulation for $t<3000$ with $\sigma=\beta=0.01$. The color code is the same as that used in {\bf b} and {\bf c}.}\label{fig:Fig4}
\end{figure}

\noindent{\bf Revisiting molecular chaos}
\parskip = 0pt plus 1pt

To evaluate the effect of clustering on $\mu$, we start by categorizing different types of binary collisions. In the dilute limit, binary collisions are considered free from interference from other particles. Consequently, the initial contact between an isolated pair (Particle 1 and 2) is possible only when $\vec{r}_{12}\cdot \vec{n}_{21}>0$, a case that we shall refer to as ``normal'' collisions. Here, $\vec{n}_{21} = \vec{n}_2 - \vec{n}_1$ indicates the relative orientation of the two particles. Normal collisions are fully described by two parameters: the impact parameter $b\equiv |\vec{r}_{12} \times \vec{n}_{21}|/|\vec{r}_{12}| \in[0,1]$ and the collision angle $\Delta \equiv \cos^{-1}(\vec{n}_{1}\cdot\vec{n}_{2})\in[0,\pi]$ (Fig.\,\ref{fig:Fig4}a) \cite{Lam_selfpropelled_2015,Hanke_2013}. At high densities inside clusters, however, the direction of particle motion can deviate substantially from particle orientation $\vec{n}$ due to interactions among multiple particles (Fig.\,\ref{fig:Fig4}c). Indeed, we observe increasingly frequent collisions with $\vec{r}_{12}\cdot\vec{n}_{21}<0$ in clusters at large $\rho_c$ (Fig.\,\ref{fig:Fig4}e), which are ``abnormal'' in comparison with the collisions in the dilute limit.

To characterize abnormal collisions, we replace $b$ by a new parameter $\gamma_{1}\equiv \cos^{-1}[(\vec{r}_{12}\cdot \vec{n}_{21})/|\vec{r}_{21}|]$,  i.e., the angle between $\vec{r}_{12}$ and $\vec{n}_{21}$ (Fig.\,\ref{fig:Fig4}a). For normal collisions, $\gamma_1$ and $b$ are equivalent through $\gamma_{1}=\sin^{-1}b\in[0,\pi/2]$. For abnormal collisions (Fig.\,\ref{fig:Fig4}c), $b$ is ill-defined while $\gamma_1 > \pi/2$ remains valid. By introducing a second parameter $\gamma_2$ equivalent to $\Delta$ through $\gamma_2 = (\pi-\Delta)/2$ (Fig.\,\ref{fig:Fig4}a), normal and abnormal collisions can be further divided into two sub-categories: aligned (anti-aligned) collisions where the tangential motions of the two colliding particles are in the same (opposite) direction. Thus, all binary collisions are divided into four categories (Fig.~\ref{fig:Fig4}b,c) where $\vec{p}\cdot\delta\vec{p}$ can be analytically calculated as (SI Sec.\, E): (1) $4\sin\gamma_2[\sin(\gamma_2+\beta)-\sin\gamma_2]$ for normal aligned collisions with $0<\gamma_1<\gamma_2<\pi/2$; (2) $-8\sin^{2}\gamma_2\sin^{2}\frac{\beta}{2}$ for normal anti-aligned collisions with $0<\gamma_2<\gamma_1<\pi/2$; (3) $4\sin\gamma_2[\sin(\gamma_2-\beta)-\sin\gamma_2]$ for abnormal aligned collisions with $0<(\pi-\gamma_1)<\gamma_2<\pi/2$; and (4) $-8\sin^{2}\gamma_2\sin^{2}\frac{\beta}{2}$ for abnormal anti-aligned with $0<\gamma_2<(\pi-\gamma_1)<\pi/2$.

Within each category, our simulation for $t<3000$ with $\sigma=\beta=0.01$ shows that the average momentum change $\mu$ is approximately constant independent of $\rho_c(t)$ with $\mu_{1} \approx 1.1\times 10^{-2}$, $\mu_{2} =\mu_{4} \approx -10^{-4}$, and $\mu_{3} \approx -0.9\times 10^{-2}$ (Fig.\,\ref{fig:Fig4}d), where $\mu_i$ is obtained by averaging $\vec{p}\cdot\delta\vec{p}$ over all observed collisions in Category ($i$) with $i=1,2,3,4$. The results are in good agreement with the calculation assuming molecular chaos within each category: averaging the analytic expressions of $\vec{p}\cdot\delta\vec{p}$ in the previous paragraph over the uniform distribution of $\gamma_1$ and $\gamma_2$ gives $\mu_1=-\mu_3 = 4\beta/\pi = 1.3\times 10^{-2}$ and $\mu_2=\mu_4 = -(1-4/\pi^2)\beta^2 = -6\times 10^{-5}$ to the leading order in $\beta$ (SI Sec.\,E). Thus, the temporal evolution of the total momentum change averaged over all four categories $\mu(t)=\sum_{i=1}^{4} w_{i}(\rho_c)\mu_{i}$ arises almost entirely from the $\rho_c(t)$ dependence of the relative frequency $w_{i}$ of the four categories (Fig.\,\ref{fig:Fig4}e). 

To predict $\mu(t)$, we propose a kinetic model with two simple assumptions on $w_{i}$. First, we assume that $w_1=0.5$ is a constant, which naturally reduces to molecular chaos if there are no abnormal collisions. Since abnormal collisions must arise from many-body interactions, we further assume that $w_3$ is $1/2$ for particles in clusters (abnormal and normal collisions are equally probable) and 0 for particles outside clusters. For a system with with a fraction of particles in clusters $\rho_c$, $w_3=\rho_c/2$.

\begin{figure}[t]
	\includegraphics[width=0.46\textwidth]{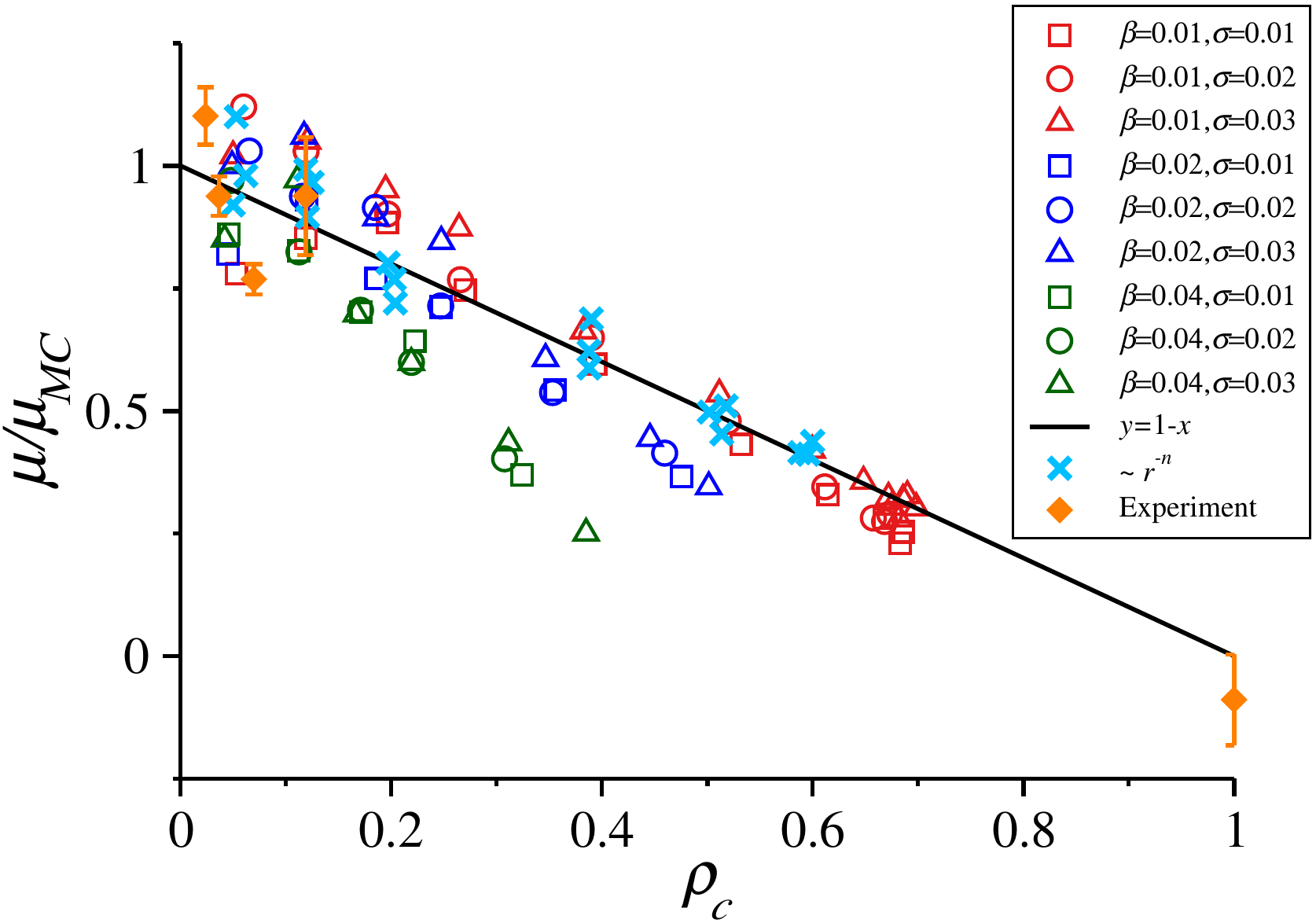}
    \caption{{\bf Clustering-dependent transport coefficient.} The transport coefficient $\mu$ versus the fraction of particles in clusters $\rho_c$, obtained from simulations with different $\beta$ and $\sigma$ (open symbols and blue crosses) and experiments of different area fractions $\phi$ (solid diamond). The solid line is the theoretical prediction (Eq.~\ref{improved-MC}). $\mu$ is normalized by the prediction of the original molecular chaos hypothesis, $\mu_{MC}$.}\label{fig:Fig5}
\end{figure}

Applying these two assumptions along with molecular chaos within each category, we reach
\begin{equation}
\mu = w_{1}\mu_{1}+w_{3}\mu_{3}=(1-\rho_c)\mu_{MC}
\label{improved-MC}
\end{equation} 
to the linear order of $\beta$. The contributions from $\mu_2$ and $\mu_4$ are neglected as they are second order in $\beta$. Equation~\ref{improved-MC} shows that clustering determines the deviation away from the original molecular chaos prediction.

To verify the prediction, we obtain $\mu$ from both experiments and simulations via Eq.\,\ref{kinetic-theory} at various $t$ with $M<0.3$. Since experiments are conducted in a circular arena, an extra term is introduced at the right hand side of Eq.~\ref{kinetic-theory} to account for the boundary contribution to the change in $M$. The time average of $\mu(t)$ over the steady state is taken to reduce the noise due to the small experimental system size. The detail of our experimental analysis is given in SI Sec.\,F. Figure~\ref{fig:Fig5} shows $\mu/\mu_{MC}$ versus $\rho_c$. For different sets of $\{\beta,\sigma\}$, system dynamics from simulation follow Eq.\,\ref{improved-MC} well. Since only steady states with limited parameter choices are available, experiments (solid diamonds) show two distinctive behaviors both in good agreement with the model prediction: $\mu>0$ favoring alignment with little clustering at intermediate $\phi$ between $0.3$ to $0.45$, whereas $\mu \approx 0$ at high $\phi=0.7$ where almost all particles are in clusters with $\rho_c \approx 1$. Simulations with soft repulsive power-law potentials (crosses) also lead to the same relation between $\mu/\mu_{MC}$ and $\rho_c$ (Fig.~\ref{fig:Fig5}), where we use $\mu(\rho_c \to 0)$ from simulation as $\mu_{MC}$ since its value depends on the specific form of the potential and is generally not analytically available. The quantitative agreement between $\mu/\mu_{MC}$ and $1-\rho_c$ at different times is also shown in Fig.~\ref{fig:Fig3}b. 

\parskip = 5pt plus 1pt

\noindent{\bf Optimal growth rate of flocking}
\parskip = 0pt plus 1pt

Lastly, we study the $\phi$ dependence of $dM/dt$ for a given $\{\sigma,\beta\}$. In the dilute limit, molecular chaos applies so that $\mu=\mu_{MC}$ is constant, whereas the collision rate $\lambda \sim \phi$ according to the ideal gas assumption. Therefore, $dM/dt$ increases as $\phi$ increases according to Eq.\,\ref{kinetic-theory}. When $\phi$ increases above the characteristic density of MIPS, the suppression of $\mu$ by clustering dictated by Eq.\,\ref{improved-MC} leads to a nonmonotonic trend of $dM/dt$ versus $\phi$.

\begin{figure}
	\includegraphics[width=0.46\textwidth]{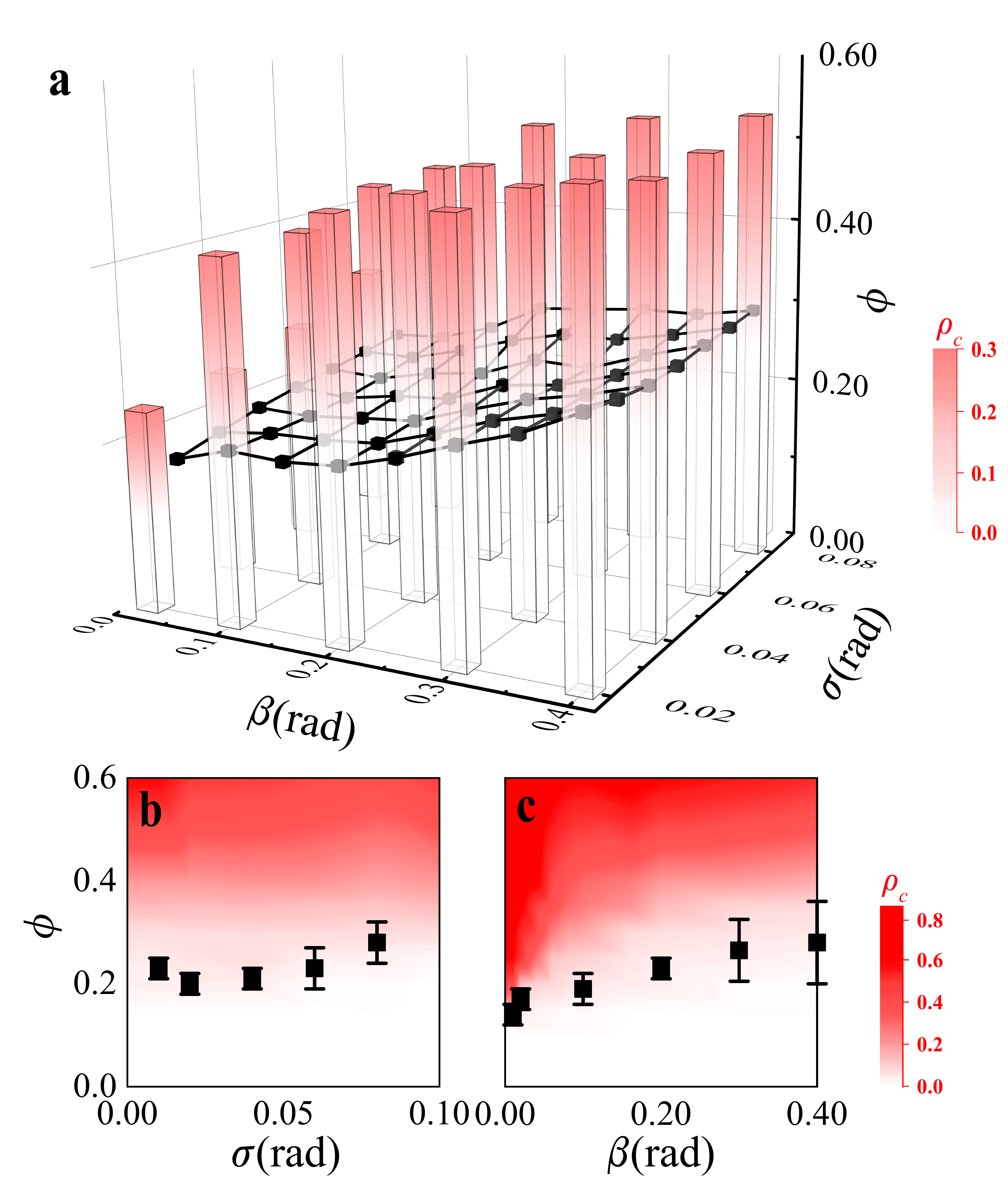}
	\caption{{\bf Optimal growth rate of flocking.} The optimal area fraction with the fastest growth of flocking $\phi^*$ (black squares) in the three dimensional (3D) phase diagram of MIPS defined by $\beta$, $\sigma$ and $\phi$ ({\bf a}), and in a two dimensional (2D) phase diagram at fixed $\beta=0.2$ ({\bf b}) and at fixed $\sigma=0.01$ ({\bf c}).}\label{fig:Fig6}
\end{figure}

We measure the time for flocking to be fully developed at given $\{\sigma,\beta\}$, $t_{f}$, as a function of $\phi$, from which we find the optimal $\phi = \phi^*$ with the smallest $t_{f}$ (Fig.\,S7 and SI Sec.\,H). Interestingly, $\phi^*$ appears right below the phase boundary of MIPS (Fig.\,\ref{fig:Fig6}), which provides the maximum possible particle interactions promoting alignment while avoiding the suppression of $\mu$ due to the effect of particle clustering. In other words, the growth of flocking is fastest for a system in the marginal state barely escaping the fate of MIPS. Thus, the phase boundary of MIPS also marks the ridge in kinetics where the growth rate of flocking is highest.

\parskip = 5pt plus 1pt

\noindent {\bf Discussion}
\parskip = 0pt plus 1pt

Our system differs from passive systems due to the inclusion of self-propulsion, a new degree of freedom that can be decoupled from translational velocity during collision. As a result, there are three time scales at play: the time scale of particle reorientation $\tau_R\approx(\beta+\sigma^2)^{-1}$, the mean free flight time $\tau_f$, and the time scale for particles to be pushed apart due to the short-ranged repulsion during collision $\tau_C \approx 1$. While $\tau_f \sim \phi^{-1}$ in the dilute limit can be extended to relatively high densities for passive systems, it is invalid for even moderately dense active systems due to the formation of densely packed clusters where $\tau_f$ becomes really small. Two implications thus follows. First, it is necessary to consider binary collision at each time step, instead of binary scattering events that are previously studied as the period between the initial contact and the eventual escape of two particles \cite{Lam_selfpropelled_2015}. Such events are well-defined only when the events can proceed without interruption with $\tau_R < \tau_f$. For a typical active system with $\beta, \sigma \ll 1$, even at moderate density (e.g. $\phi=0.3$), scattering events are frequently interrupted and therefore ill-defined. Second, the assumption of molecular chaos is valid only when $\tau_C \ll \tau_f$, which is satisfied for passive systems at relatively high densities. But the condition cannot be met in even moderately dense active systems due to the formation of clusters, which in principle requires the more general Bogoliubov-Born-Green-Kirkwood-Yvon (BBGKY) approach.

The simplicity of our system allows us to quantify the crucial effect of many-body interactions and develop a simple kinetic model that avoids the complex BBGKY hierarchy. Particularly, we show that many-body interactions can be treated implicitly by considering four different categories of binary collisions, which leads to a clustering correction to the molecular chaos prediction (Eq.\,\ref{improved-MC}). The identification of the four categories of binary collisions is generically applicable for active systems. Our results with different types of repulsive potentials further suggests that the hypothesis of molecular chaos within each category could also be a general feature. Thus, our study provides not only important insights into the dynamics of natural and engineering active systems showing collision avoidance but also a promising theoretical approach for understanding dense active systems.

{\bf Acknowledgements}
We thank L. Gordillo and A. Khlyustova for the help with experiments, and H. Chat{\' e}, O. Dauchot, L. S. Luo, and M. Stynes for constructive discussions. This research was supported by the National Natural Science Foundation of China 11974038, 11750110409, and U2230402 and by US National Science Foundation CBET-1702352 and CBET-2028652. We also acknowledge computational support from the Beijing Computational Science Research Center and Minnesota Supercomputing Institute of University of Minnesota.

{\bf Author contributions}
K.J.W., X.C., and X.X. designed research; K.J.W., D.G., T.-P.S., and J.K. performed experiments; L.C., K.J.W., P.L. and B.Z. conducted numerical simulations; L.C., K.J.W., P.L., Z.T., X.C., and X.X. analyzed data; X.C. and X.X. supervised the project and co-wrote the paper. All authors discussed and commented on the manuscript.

{\bf Additional information}
Correspondence and requests for materials should be addressed to Xiang Cheng (xcheng@umn.edu), and Xinliang Xu (xinliang@csrc.ac.cn).

{\bf Competing financial interests}
The authors declare no conflict of interest.

{\bf Data availability}
All data that support the plots within this paper and other findings of this study are available from the corresponding authors upon request.

{\bf Code availability}
All computer code used for this study is available from the corresponding authors upon request.

\end{document}